\documentclass{article}
\usepackage{dcase2023,amsmath,url,times,booktabs,tabularx,amsmath,amssymb}
\usepackage{array,multirow}
\usepackage{arydshln}
\usepackage{multicol}
\usepackage{footnote}
\usepackage{float}
\usepackage{cite}
\usepackage{enumitem}
\usepackage{color,soul}
\usepackage{subcaption}

\usepackage[pdftex]{graphicx}

\newcolumntype{C}[1]{>{\centering\arraybackslash}p{#1}}
\newcolumntype{L}[1]{>{\raggedright\arraybackslash}p{#1}}
\newcolumntype{R}[1]{>{\raggedleft\arraybackslash}p{#1}}

\makeatother
\newcommand{\vect}[1]{{\mbox{\boldmath $#1$}}}
\newcommand{\T}[0]{\mathsf{T}}

\title{DESCRIPTION AND DISCUSSION ON DCASE 2025 CHALLENGE TASK 2: FIRST-SHOT UNSUPERVISED ANOMALOUS SOUND DETECTION FOR MACHINE CONDITION MONITORING}

\name{
Tomoya Nishida$^{1}$, Noboru Harada$^{2}$, Daisuke Niizumi$^{2}$, Davide Albertini$^{3}$, Roberto Sannino$^{3}$,
}
\secondlinename{
Simone Pradolini$^{3}$, Filippo Augusti$^{3}$, Keisuke Imoto$^{4}$, Kota Dohi$^{1}$, Harsh Purohit$^{1}$,
}
\thirdlinename{
Takashi Endo$^{1}$, and Yohei Kawaguchi$^{1}$
}
\address{
$^1$ Hitachi, Ltd., Japan, \url{tomoya.nishida.ax@hitachi.com}\\
$^2$ NTT, Inc., Japan, \url{harada.noboru@ntt.com}\\
$^3$ STMicroelectronics, Switzerland, \url{}\\
$^4$ Kyoto University, Japan, \url{keisuke.imoto@ieee.org}\\
}

\begin{document}

\ninept
\maketitle

\begin{sloppy}

\begin{abstract}
This paper introduces the task description for the Detection and Classification of Acoustic Scenes and Events (DCASE) 2025 Challenge Task 2, titled “First-shot unsupervised anomalous sound detection (ASD) for machine condition monitoring.” 
Building on the DCASE 2024 Challenge Task 2, this task is structured as a first-shot problem within a domain generalization framework.
The primary objective of the first-shot approach is to facilitate the rapid deployment of ASD systems for new machine types without requiring machine-specific hyperparameter tunings.
For DCASE 2025 Challenge Task 2, sounds from previously unseen machine types have been collected and provided as the evaluation dataset.
We received 119 submissions from 35 teams, and an analysis of these submissions has been made in this paper.
Analysis showed that various approaches can all be competitive, such as fine-tuning pre-trained models, using frozen pre-trained models, and training small models from scratch, when combined with appropriate cost functions, anomaly score normalization, and use of clean machine and noise sounds.
\end{abstract}


\section{Introduction}
\label{sec:intro}
\vspace{-6pt}
Anomalous sound detection (ASD)~\cite{koizumi2017neyman, kawaguchi2017how, koizumi2019neyman, kawaguchi2019anomaly, koizumi2019batch, suefusa2020anomalous, purohit2020deep} involves determining whether the sound emitted from a target machine is normal or anomalous.
This capability plays a crucial role in automating the detection of mechanical failures, which is essential in the era of the fourth industrial revolution and AI-driven factory automation.

One of the key challenges in developing ASD systems lies in the scarcity and limited diversity of anomalous samples available for training.
To address this, the first ASD task was introduced in the DCASE Challenge 2020 Task 2~\cite{Koizumi2020dcase}, focusing on “unsupervised ASD~(UASD),” which aimed to detect unknown anomalous sounds using only normal sound samples for training. 
Building on this, subsequent challenges in 2021 and 2022~\cite{Kawaguchi2021, Dohi2022DCASE} tackled the issue of domain shifts to enable broader application of ASD systems. 
Domain shifts refer to discrepancies between data from the source and target domains, arising due to variations in machine operational conditions or environmental noise.

The DCASE 2023/2024 Task 2 (“first-shot” UASD)~\cite{Dohi2023DCASE, Nishida2024DCASE} targeted a realistic setting where systems must detect anomalies for entirely novel machine types without access to similar-type data for training or hyperparameter tuning.
This reflects rapid-deployment scenarios in which collecting diverse training or test data—especially anomalous samples—is infeasible, and therefore manual test-driven tuning is unrealistic.
Accordingly, the evaluation data comprise machine types absent from the development set to enforce this constraint.

The DCASE2025 Challenge Task 2 maintains the previous task setting as a first-shot problem under domain generalization conditions, using newly recorded machine sound data as the evaluation dataset.
In addition, there are several modifications:
We provide additional supplementary data for each machine, including clean machine recordings or noise samples, which can optionally be used to enhance ASD performance in noisy environments.
Also, participants are asked to provide the computational complexity of their solutions. 
Although this score is not used for the official rankings, it helps clarify the balance between model complexity and performance—a key factor for lightweight ASD applications on edge devices.
In this paper, we provide explanations on this task and discuss the challenge results.

\vspace{-6pt}
\section{First-shot Unsupervised Anomalous Sound Detection under Domain Shifted Conditions} 
\label{sec:uasd}
\vspace{-6pt}
Consider an audio clip $\vect{x}$, which contains sounds produced by a machine.
The objective of the ASD task is to classify the machine as either normal or anomalous by calculating an anomaly score $\mathcal{A}_{\theta}(\vect{x})$ using an anomaly score calculator $\mathcal{A}$ with parameters $\theta$. 
The input of $\mathcal{A}$ can be the audio clip $\vect{x}$ with or without additional information such as labels indicating the operation condition of the machine.
The machine is then determined to be anomalous when $\mathcal{A}_{\theta}(\vect{x})$ exceeds a pre-defined threshold $\phi$ as
\begin{equation}
\mbox{Decision} = \left\{
\begin{array}{ll}
\mbox{Anomaly} & (\mathcal{A}_{\theta}(\vect{x}) > \phi)\\
\mbox{Normal} & (\mbox{otherwise}).
\end{array}
\right.
\label{eq:det}
\end{equation}
The primary difficulty in this task is to train the anomaly score calculator with only normal sounds (UASD). 
The DCASE 2020 Challenge Task 2~\cite{Koizumi2020dcase} was designed to address this issue, and all the following tasks stand on this UASD setting.

Addressing the domain-shift problem is also essential for the practical implementation of ASD systems.
Domain shifts refer to variations in conditions between training and testing phases, which alter the distribution of the observed sound data.
These variations can result from differences in operating speed, machine load, heating temperature, microphone arrangement, environmental noise, and other factors.
Two domains are defined: the \textbf{source domain}, representing the original condition with sufficient training data, and the \textbf{target domain}, representing another condition where only limited samples are available.
This year's task follows the 2022 to 2024 Task 2~\cite{Dohi2022DCASE, Dohi2023DCASE, Nishida2024DCASE} setting, where the domain information is assumed to be unknown in the test phase and anomalies from both domains have to be detected with a single threshold.
In this case, domain generalization is required to achieve good performance.

To further pursue the rapid development of ASD systems in real-world scenarios, solving ASD (a) against completely novel machine types (b) with only one section of training data (c) without handcrafted tunings that depend on test data, are highly important.
This is because in real-world scenarios, customers may only possess a single novel machine, and collecting test data-especially the anomalous samples-for handcrafted tuning may be infeasible.
This problem setting was named as the “first-shot problem”, and the 2023 and 2024 Task 2~\cite{Dohi2023DCASE, Nishida2024DCASE} was organized based on this problem setting.
The first-shot problem was implemented by introducing two key features to the dataset:
(i) The development and evaluation datasets consist of entirely different sets of machine types, and
(ii) Each machine type in the dataset contains only a single section.
Note that until 2022 Task 2, the provided dataset included multiple sections for each machine type, and the development and evaluation datasets sharing the same machine types.

The DCASE2025 Challenge Task 2 retains the previous task setting as a first-shot problem under domain generalization conditions, while introducing several modifications.
First, we have provided additional supplementary data, including clean machine recordings and noise samples. 
These resources may reflect practical scenarios—such as collecting clean machine data when a factory is idle or gathering noise recordings when the machine is not running.
Participants are free to incorporate these additional sources to enhance the accuracy of their models.
Second, although large-scale models—such as pretrained networks and ensembles—have become increasingly popular in this task, lightweight models capable of running on edge devices also remain an important area of research.
To acknowledge this, participants were optionally asked to report the computational complexity of their solutions in terms of Multiply-Accumulate Operations (MACs).
Although this metric does not affect the official rankings, it provides valuable insight into the balance between model complexity and performance.

\section{Task Setup} 
\label{sec:task}
\vspace{-6pt}

\subsection{Dataset} 
\label{sec:dataset}
\vspace{-5pt}
The dataset for this task is divided into three categories: the \textbf{development dataset}, the \textbf{additional training dataset}, and the \textbf{evaluation dataset}. 
The development dataset contains seven machine types, while the additional training and evaluation datasets include nine machine types, with each machine type consisting of a single section.
A \textbf{machine type} refers to the category of machines, such as fans or gearboxes, and a \textbf{section} represents a subset or the entirety of the data associated with each machine type.

All recording are single-channel, lasting 6 to 10 seconds, and have a sampling rate of 16 kHz. 
The machine sounds recorded at laboratories were mixed with environmental noise recorded at factories and in the suburbs to create each sample in the dataset.
For further details of the recording process, please refer to the papers on ToyADMOS2~\cite{harada2021toyadmos2} and MIMII DG~\cite{Dohi2022}.

The \textbf{development dataset} provides seven machine types (fan, gearbox, bearing, slide rail, valve, ToyCar, ToyTrain), and each machine type has one section that contains a complete set of the training and test data.
Each section contains
(i) 990 normal clips from a source domain for training, 
(ii) 10 normal clips from a target domain for training,
(iii) 100 clips of supplementary sound data containing either clean normal machine sounds in the source domain or noise-only sounds, and
(iv) 100 normal clips and 100 anomalous clips from both domains for the test.
To assist participants, domain information (source/target) was included in the test data. 
For four machine types (fan, gearbox, valve, and ToyCar) details regarding operational or environmental conditions were provided in the file names and attribute CSV files. 
However, for the remaining three machine types, these attributes were not disclosed.

The \textbf{additional training dataset} provides novel nine machine types (AutoTrash, HomeCamera, ToyPet, ToyRCCar, BandSealer, Polisher, ScrewFeeder, CoffeeGrinder).
Each section consists of 
(i) 990 normal clips in a source domain for training,
(ii) 10 normal clips in a target domain for training. and
(iii) 100 clips of supplementary sound data containing either clean normal machine sounds in the source domain or noise-only sounds.
For five machine types (HomeCamera, ToyRCCar, BandSealer, and CoffeeGrinder), attributes were provided in this dataset.
For the other four machine types, attributes were concealed.
The \textbf{evaluation dataset} provides the test clips that correspond to the additional training dataset, e.g. data of the same machine types as the additional training dataset. 
Each section consists of 200 test clips, none of which have a condition label (i.e., normal or anomaly), domain information, or attribute information. 
Participants are required to train a model for each new machine type using only a single section per machine type.

\subsection{Evaluation metrics} 
\label{sec:metrics}
\vspace{-5pt}
To assess overall detection performance, we employed the area under the receiver operating characteristic curve (AUC). 
Additionally, we used the partial AUC~(pAUC) to evaluate performance in a low false-positive rate range $[0, p]$, where we set $p = 0.1$.
To evaluate each system under the domain generalization setting, we compute the AUC for each domain and pAUC for each section as
\begin{equation}
	{\rm AUC}_{m, n, d} = \frac{1}{N^{-}_{d}N^{+}_{n}} \sum_{i=1}^{N^{-}_{d}} \sum_{j=1}^{N^{+}_{n}}
	\mathcal{H} (\mathcal{A}_{\theta} (x_{j}^{+}) - \mathcal{A}_{\theta} (x_{i}^{-})),
\end{equation}
\begin{equation}\text{\scalebox{0.93}{$
	{\rm pAUC}_{m, n} = \frac{1}{\lfloor p N^{-}_{n} \rfloor N^{+}_{n}} \sum_{i=1}^{\lfloor p N^{-}_{n} \rfloor N^{+}_{n}} \sum_{j=1}^{N^{+}_{n}}
	\mathcal{H} (\mathcal{A}_{\theta} (x_{j}^{+}) - \mathcal{A}_{\theta} (x_{i}^{-})),
$}}
\end{equation}
where $m$ and $n$ represent the index of a machine type and a section respectively,
$d \in \{ {\rm source}, {\rm target} \}$ represents a domain,
$\lfloor \cdot \rfloor$ is the flooring function,
and $\mathcal{H} (y)$ returns 1 when $y > 0$ and 0 otherwise.
Here, $\{x^{-}_{i}\}_{i=1}^{N^{-}_{d}}$ are the normal test clips in domain $d$ in section $n$ of machine type $m$ and $\{x_{j}^{+}\}_{j=1}^{N^{+}_{n}}$ are all the anomalous test clips in section $n$ of machine type $m$.
$N^{-}_{d}, N^{-}_{n}, N^{+}_{n}$ represent the number of normal test clips in domain $d$, normal test clips in section $n$, and anomalous test clips in section $n$, respectively.

The official score $\Omega$ is given by the harmonic mean of the AUC and pAUC scores overall machine types and sections:
\begin{eqnarray}
\Omega &=& h \left\{ {\rm AUC}_{m, n, d}, \ {\rm pAUC}_{m, n} \quad | \quad \right. \nonumber \\
&& \left. m \in \mathcal{M}, \  n \in \mathcal{S}(m), \ d \in \{ {\rm source}, {\rm target} \} \right\},
\end{eqnarray}
where $h\left\{\cdot\right\}$ represents the harmonic mean, $\mathcal{M}$ is the set of given machine types, and $\mathcal{S}(m)$ represents the set of sections for machine type $m$.
Specifically, $\mathcal{S}(m)=\{00\}$ for the dataset in 2024-2025.

Additionally, although not included in the official rankings, participants were optionally asked to provide information on the computational complexity of their models in terms of MAC operations. It was recommended that this be calculated using the open-source implementation available in \cite{thop}.

\subsection{Baseline systems and results}
\label{sec:baseline}
\vspace{-5pt}
The task organizers offer a baseline system using Autoencoders (AEs) with two operating modes, identical to the 2023 Task 2 baseline.
While both modes use Autoencoders for training, they differ in anomaly score computation.
This paper presents the system and its detection performance; details can be found in \cite{Harada2023}.

\subsubsection{Autoencoder training}
\vspace{-5pt}
The AE is trained for both operating modes using log-mel-spectrograms of training sound clips $X = [X_1, \dots, X_T]$, where $X_t \in \mathbb{R}^F$ for $t=1,\dots,T$ represents frame-wise feature vectors at frame $t$, where $F=128$ and $T$ is the number of mel-filters and time-frames, respectively.
For input, $P=5$ consecutive frames are concatenated as $\psi_t = [X_t^\T, \dots, X_{t + P - 1}^\T]^\T \in \mathbb{R}^{D}$ for each $t$, with $D=P \times F = 640$.
Model parameters are trained by minimizing the mean squared error~(MSE) between the input $\psi_t$ and the reconstructed output $r_\theta (\psi_t)$ for all inputs from the training data.

\subsubsection{Simple Autoencoder mode}
\vspace{-5pt}
This mode uses the mean MSE of all features derived from the given sound clip as its anomaly score, e.g.,
\begin{equation}
A_{\theta}(X) = \frac{1}{DK} \sum_{k = 1}^K \| \psi_k - r_{\theta}(\psi_k) \|_{2}^{2},
\end{equation}
where $K=T-P+1$, and $\| \cdot \|_2$ represents $\ell_2$ norm.

\subsubsection{Selective Mahalanobis mode}
\vspace{-5pt}
In this mode, the Mahalanobis distance between the system input and reconstructed feature is used to compute the anomaly score. 
The anomaly score is defined as
\begin{align}
&\text{\scalebox{0.95}{$A_{\theta}(X) = \frac{1}{DK} \sum_{k = 1}^K \min\{ D_s (\psi_k, r_{\theta}(\psi_k)), D_t (\psi_k, r_{\theta}(\psi_k))\}$}},\\
&D_s(\cdot) = \textrm{Mahalanobis}(\psi_k, r_{\theta}(\psi_k), \Sigma_s^{-1}), \\
&D_t(\cdot) = \textrm{Mahalanobis}(\psi_k, r_{\theta}(\psi_k), \Sigma_t^{-1}),
\end{align}
where $\Sigma_s^{-1}$ and $\Sigma_t^{-1}$ are the covariance matrices of $r_{\theta}(\psi_k) - \psi_k$ for the source and target domain data of each machine type, respectively.

\subsubsection{Results}
\label{sec:results}
\setlength{\tabcolsep}{1mm}

\begin{table}[t]
\begin{center}
\vspace{-10pt}
\caption{Baseline results for development dataset.}
\label{tab:baseline_results}
\scriptsize
\begin{tabular}{@{}c c c c p{1pt} c c@{}}
\hline
\ \\[-6.5pt]
Machine type &
Mode &
\multicolumn{2}{c}{AUC [\%]} &&
\multicolumn{2}{c}{pAUC [\%]} \\
\cline{3-4} \cline{6-7}
\ \\[-6.5pt]
& & 
\multicolumn{1}{c}{Source} &
\multicolumn{1}{c}{Target} && \\
\hline
\ \\[-6.5pt]
  ToyCar
  &	MSE & $71.05 \pm 0.50$ & $53.32 \pm 0.56$ && $49.79 \pm 0.49$\\
  
  &	MAHALA & $73.17 \pm 0.39$ & $50.91 \pm 0.85$ && $49.05 \pm 0.05$\\
  \hline
  
  ToyTrain
  &	MSE & $61.76 \pm 0.74$ & $56.46 \pm 0.47$ && $50.19 \pm 0.25$\\
  
  &   MAHALA & $50.87 \pm 2.88$ & $46.15 \pm 1.77$ && $48.32 \pm 0.05$\\
  \hline
  
  bearing
  &	MSE & $66.53 \pm 2.63$ & $53.15 \pm 1.99$ && $61.12 \pm 0.59$\\
  
  &	MAHALA & $63.63 \pm 1.15$ & $59.03 \pm 1.79$ && $61.86 \pm 0.36$\\
  \hline
  
  fan
  &	MSE & $70.96 \pm 0.94$ & $38.75 \pm 0.74$ && $49.46 \pm 0.53$\\
  
  &	MAHALA & $77.99 \pm 0.23$ & $38.56 \pm 0.58$ && $50.82 \pm 0.06$\\
  \hline
  
  gearbox
  &	MSE & $64.80 \pm 1.48$ & $50.49 \pm 1.22$ && $52.49 \pm 0.37$\\
  
  &	MAHALA & $73.26 \pm 0.78$ & $51.61 \pm 0.52$ && $55.07 \pm 0.47$\\
  \hline
  
  slider
  &	MSE & $70.10 \pm 1.01$ & $48.77 \pm 1.07$ && $52.32 \pm 0.36$\\
  
  &	MAHALA & $73.79 \pm 1.95$ & $50.27 \pm 1.15$ && $53.61 \pm 0.26$\\
  \hline
  
  valve
  &	MSE & $63.53 \pm 2.90$ & $67.18 \pm 1.75$ && $57.35 \pm 1.96$\\
  
  &	MAHALA & $56.22 \pm 2.22$ & $61.00 \pm 2.98$ && $52.53 \pm 1.32$\\
  \hline
\end{tabular}
\end{center}
\end{table}

Tables \ref{tab:baseline_results} present the AUC and pAUC results for the two baseline systems on the development dataset, with the averages and standard deviations computed from five independent trials.

\section{Challenge Results}
\begin{figure*}[t]
	\begin{center}
        \includegraphics[width=1.0\hsize,clip]{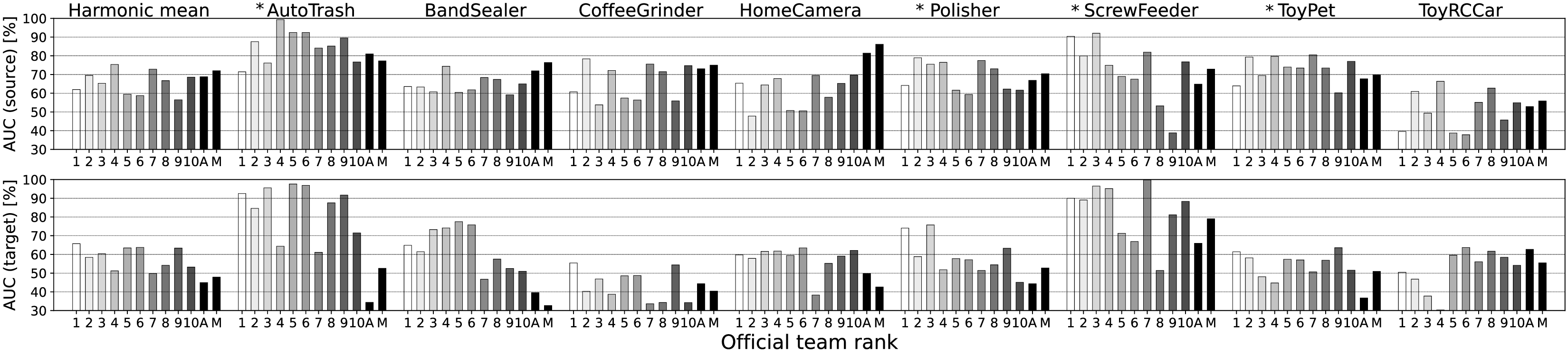}
        \vspace{-13pt}
	\caption{Evaluation results of top 10 teams in ranking. Average source (top) and target-domain AUC (bottom) for each machine type~(``*" indicates that attributes are hidden.). Labels ``A'' and ``M'' on denote simple Autoencoder mode and selective Mahalanobis mode, respectively.}
	\label{fig:aucs}
	\end{center}
\end{figure*}
\begin{figure}[t]
    \center
    \includegraphics[width=0.63\linewidth,clip]{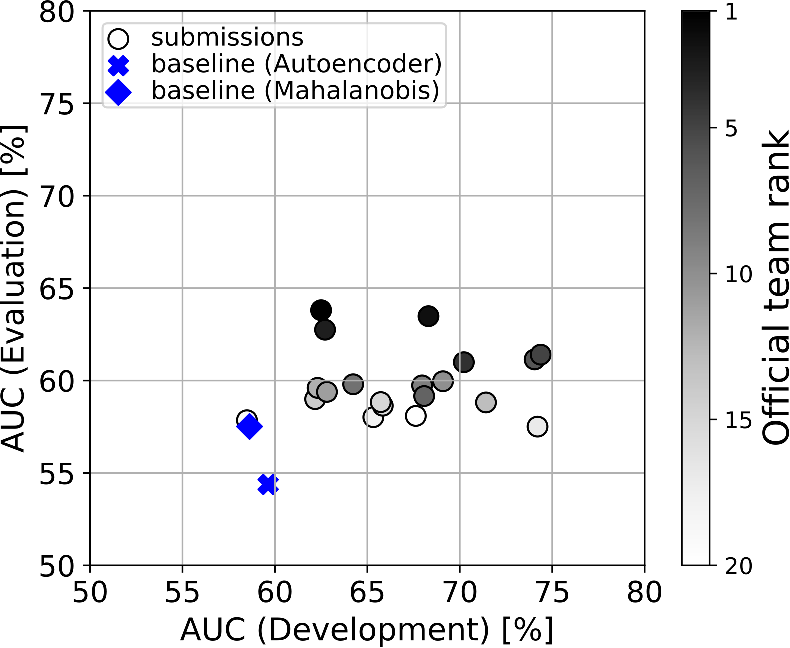}
    \caption{Comparison of harmonic mean of AUC for development and evaluation dataset across teams.}
    \vspace{7pt}
    \label{fig:dev_vs_eval}
\end{figure}
\begin{figure}[t]
    \center
    \includegraphics[width=0.63\linewidth,clip]{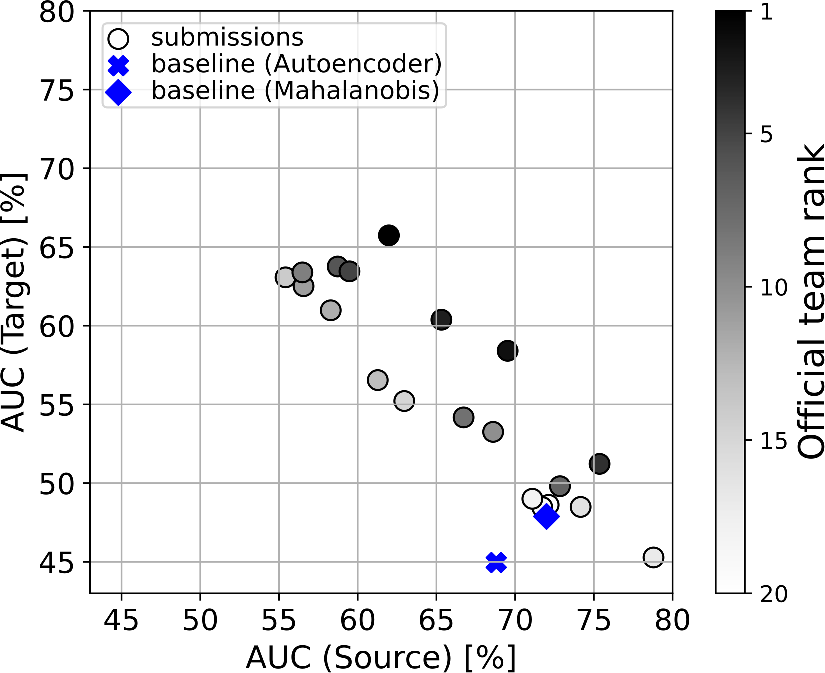}
    \caption{Comparison of harmonic mean of AUC for source and target domain in evaluation dataset across teams.}
    \label{fig:source_vs_target}
\end{figure}

\subsection{Overall results}
We received 119 submissions from 35 teams.
20 teams outperformed both baselines, which slightly increased compared to last year's task (11 out of 27 teams).
Looking at the results for each domain separately, six teams surpassed the baselines on the source-domain AUC, while 25 did so on the target-domain AUC.
Four teams achieved higher AUCs than the baselines in both domains.
This shows the difficulty of improving the performance on both the source and target domain at the same time.
Figure~\ref{fig:aucs} shows the AUC values for the top 10 teams.
In the source domain, whether each team could beat the baseline was highly machine-dependent, and many teams struggled to outperform the baseline on average.
Specifically, machines for which attribute information was available tended to perform poorly, although it is unclear whether this factor is actually relevant.
In contrast, all of the top 10 teams outperformed the baselines in the target domain in the harmonic mean.

Figure~\ref{fig:dev_vs_eval} compares the AUC values of the top 20 teams between the development and evaluation datasets.
As can be seen, achieving high AUC values in the development dataset does not indicate high AUC in the evaluation dataset.
This is a typical trend in the first-shot problem setting which started from 2023, and shows the difficulty in finding an approach robust to machine types in the absence of test data.
Thus, building an ASD system that works well for unknown machine types remains both a difficult and important challenge.
In Figure~\ref{fig:source_vs_target}, we compare the harmonic mean AUC values in the source and target domain of the evaluation dataset among the top 20 teams.
The figure shows a negative correlation between the AUC in the source and target domain.
While the top 3 teams achieved very close official scores~(within 1.0\% difference), their balance between the source and target AUC varied.
Achieving well balanced performance between the source and target domain may be important to achieve high ranks.

\subsection{Trends in model size}
\begin{figure}[t]
    \center
    \includegraphics[width=0.7\linewidth,clip]{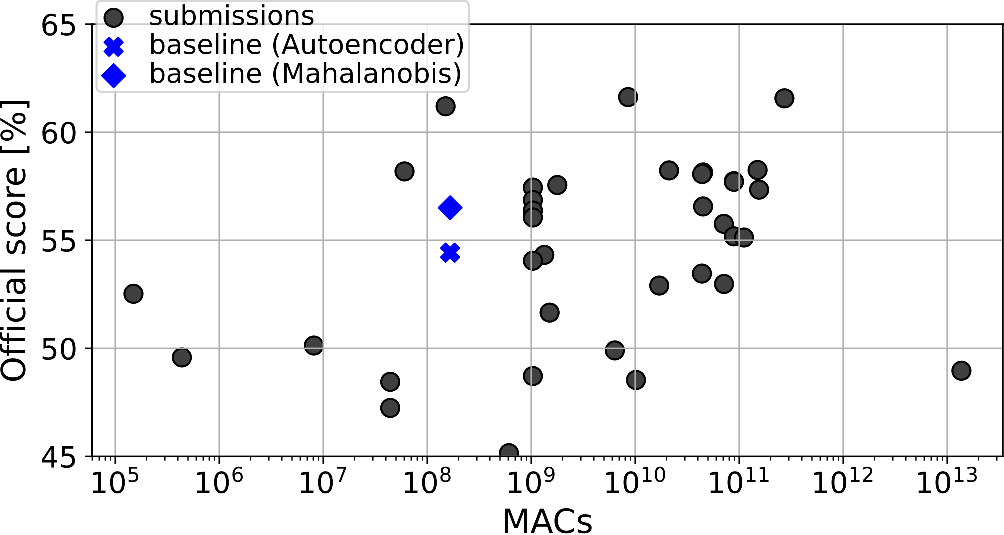}
    \caption{Comparison of officials scores of submissions against the MACs values.}
    \label{fig:macs_vs_score}
\end{figure}
We analyzed the computational-complexity trends of the submitted systems.
MACs were reported by 21 teams. 
Figure \ref{fig:macs_vs_score} plots each submission’s MAC count against its official score; if a team submitted multiple systems with identical MAC values, only the highest-ranked system was retained.

The figure shows a broad range of MAC counts across submissions.
It also highlights that larger computational budgets did not necessarily yield higher scores. 
Notably, two submissions from different teams \cite{DCASE2025-3_YangNBU2025, DCASE2025-9_ZhouXAUAT2025} achieved scores exceeding the baselines while using fewer MACs.
This shows the feasibility of computationally efficient solutions for first-shot UASD, and could be one of the future directions for research.

\subsection{New approaches seen in the top-ranked teams}
\textbf{a. Use of pretrained models}

This year, following the trend from previous years, many participants adopted pretrained models in their anomaly-detection pipelines.
Many of those teams fine-tuned them with an attribute or domain classification-based auxiliary task, such as the 1, 4, 5th ranked teams~\cite{DCASE2025-1_WangMYPS2025, DCASE2025-4_FujimuraNU2025, DCASE2025-5_JiangTHUEE2025}.
On the other hand, interestingly, several of the high-ranking teams achieved strong performance using frozen pretrained networks, exploiting intermediate-layer features along with anomaly-score normalization~\cite{wilkinghoff2025keeping}.
The 2nd~\cite{DCASE2025-2_SaengthongSCITOK2025} and 8th~\cite{DCASE2025-8_OzekiMELCO2025} place teams solely used frozen models within their submissions, while the 4th-place team~\cite{DCASE2025-4_FujimuraNU2025} ensembled frozen networks to the fine-tuned ones.
Nevertheless, teams that trained lightweight models from scratch with classification-based tasks also achieved high ranks, including the 3rd ranked team~\cite{DCASE2025-3_YangNBU2025, DCASE2025-9_ZhouXAUAT2025}, showing that pretrained models are not an absolute prerequisite for competitive performance.
Overall, diverse approaches were all competitive this year and each approach may still have room for further research.

\textbf{b. Use of supplemental data}

Participants tried utilizing the newly released supplemental clean-machine and noise recordings, applying them in two distinct ways. 
In the first way, they were used for data augmentation. 
The 1st~\cite{DCASE2025-1_WangMYPS2025} and several other top-10 teams~\cite{DCASE2025-6_ZhengSJTU-AITHU2025, DCASE2025-7_GuanHEU2025, DCASE2025-8_OzekiMELCO2025} injected the supplemental clips as an extra class in auxiliary classifiers, blended the noise signals with training samples, or leveraged them in contrastive learning~\cite{DCASE2025-9_ZhouXAUAT2025} to enrich feature space diversity. 
Conversely, the 3rd and 4th-place teams~\cite{DCASE2025-3_YangNBU2025, DCASE2025-4_FujimuraNU2025}, used the clean/noise data to build enhancement modules that extracted or denoised target machine sounds from the noisy training data, and supplied these extracted signals to their main anomaly-detection networks.

\section{Conclusion}
We presented an overview of the DCASE 2025 Challenge Task 2.
The task's aim was to develop ASD systems that work for a novel machine type with a single section for each machine type, where supplemental data such as clean machine sounds or noise-only sounds were also provided.
We discussed several new approaches seen in the challenge, such as how pretrained models were (or were not) used and the use of the newly provided supplemental data.
While we were not able to discuss all new approaches, we hope that all the technical reports will contribute to the advancements in the field of anomalous sound detection.

\clearpage
\bibliographystyle{IEEEtran}
\bibliography{refs}

\end{sloppy}
\end{document}